\documentclass[aps,nofootinbib,preprint,showpacs]{revtex4-1}
\usepackage{xcolor}
\usepackage{hyperref}
\usepackage{graphicx,subfigure}
\usepackage{amsmath,amsfonts,amssymb}

\newcommand{\f}{\frac}
\newcommand{\suml}{\sum\limits}

\newcommand{\intl}{\int\limits}

\begin{document}
\title[Zeno line]{Simple geometrical interpretation of the linear character for the Zeno-line and the rectilinear diameter}
\author{V.L. Kulinskii}
\email{kulinskij@onu.edu.ua}
\affiliation{Department for Theoretical
Physics, Odessa National University, Dvoryanskaya 2, 65026 Odessa, Ukraine}
\begin{abstract}
The unified geometrical interpretation of the linear character of the Zeno-line (unit compressibility line $Z=1$) and the rectilinear diameter is proposed. We show that recent findings about the properties of the Zeno-line and striking correlation with the rectilinear diameter line as well as other empirical relations can be naturally considered as the consequences of the projective isomorphism between the real molecular fluids and the lattice gas (Ising) model.
\end{abstract}
\maketitle
\section{Introduction}
Searching for the similarity and the unifying principles in the description of the variety of the thermodynamical properties of complex matter is the key point of statistical physics. One of the most known is the principle (theorem) of corresponding states (PCS) which goes back to van der Waals \cite{vdW_thesis}. Of course this law is approximate because of the difference in
intermolecular forces for different substances. Strictly
speaking the PCS is applicable only to the substances which
have similar interparticle potentials
\cite{book_prigozhisolut}. Moreover the PCS is pure classical
since it is based on the continuous scaling of the
characteristic of the interparticle potential. The quantum effects which lead to the discrete number of the energetic levels in the potential well lead to the deviation from this law even for heavy noble gases \cite{crit_dimers_noblepcs_physica2009}. In such a case
the search of the universal relations which applicable to broad class of substances is of great importance.
Well-known the rectilinear diameter law (RDL) \cite{crit_diam0,*crit_diam1_young_philmag1900} for the
density as the order parameter of the phase coexistence:
\begin{equation}\label{densdiam}
  n_{d} = \f{n_{liq}+n_{gas}}{2\,n_c} - 1 = A\,\left|\tau \right|+\ldots\,,\quad \tau  = \f{T-T_c}{T_c}
\end{equation}
is one of the examples. Although the relation \eqref{densdiam} is also approximate it is observed for a wide variety of fluids in a surprisingly broad temperature interval beyond the critical region where the critical fluctuations lead to the appearance of the non analytic corrections \cite{book_patpokr} (for the recent review see \cite{crit_diam_schroerweiss_statphys2008}).

Another typical phenomenological Batchinsky law was derived from the van der Waals equation\cite{eos_zenobatschinski_annphys1906}. It states that the curve defined by the equation $Z = 1$, where $Z =\f{P}{n\,T} $ is the compressibility factor, is a straight line which can be described by simple equation on $(n,T)$-plane:
\begin{equation}\label{z1}
  \f{n}{n_B}+\f{T}{T_B} = 1\,.
\end{equation}
The temperature $T_B$ corresponds to the Boyle point \cite{eos_zenobenamotz_isrchemphysj1990} and $n_B$ is the value of the density obtained by the extrapolating the coexistence curve into the low temperature region beyond triple point  \cite{eos_zenoapfelbaum_jchemp2008}. In work of Ben-Amotz and Herschbach \cite{eos_zenobenamotz_isrchemphysj1990} the line $Z = 1$ is called by the Zeno-line (ZL). They also noted the striking correlation between two remarkable linearities \eqref{densdiam}  and \eqref{z1}.

These findings have been further developed in works of Apfelbaum \textit{et al.} \cite{eos_zenoapfelbaum_jcp2004,eos_zenoapfelbaum_jchemp2006,eos_zenoapfelbaum_jchemp2008} and successfully applied for the prediction of the critical points of high-temperature metals \cite{liqmetals_zenoapfelbaum2_cpl2009}. Two main empirical facts discovered in these works are very important. The first one is the fact that the ZL tends
asymptotically to the liquid branch of binodal at low temperatures. The second fact is that the corresponding median asymptotically close to the rectilinear diameter at $T\to 0$. These facts allow to formulate the conception of the ``Triangle of Liquid-Gas States`` \cite{eos_zenoapfelbaum_jchemp2006}. The authors also claimed that the RDL can be considered as the consequence of the linear dependence of the Zeno line at low temperatures.

The aim of this work is to propose simple geometrical interpretation which naturally explains the linearities \eqref{densdiam} and \eqref{z1} as the consequence of (approximate) isomorphism of the real fluids with the lattice gas.
\section{Linearity of the Zeno-line and the density diameter}
\begin{figure}
\center
\includegraphics[scale=0.75]{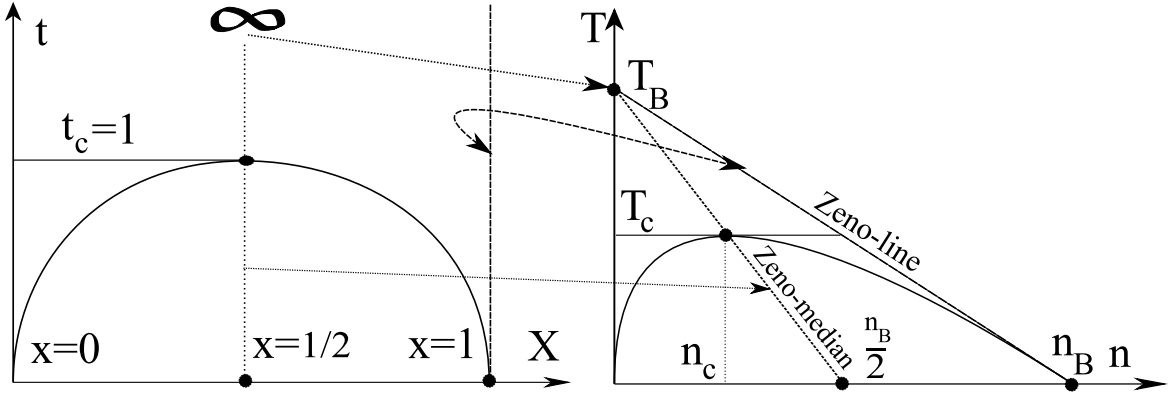}
  \caption{The geometrical correspondence between the elements of the thermodynamic phase diagram (the interaction constant is normalized $J=1$). The Zeno-line $Z=1$ and the Zeno-median are shown. The latter coincides with the diameter of the binodal .}\label{fig_zeno_tri}
\end{figure}

The geometrical fact that the ZL is the tangent to the coexistence curve allows to state that there is global isomorphism between the phase diagram of the lattice gas (LG) or equivalently the Ising model and the liquid-gas part of the diagram of the simple fluid with the coexistence region extrapolated in the region $T\to 0$. The phase diagram of the LG is remarkably symmetrical (see \ref{fig_zeno_tri}). This is the reflection of the particle-hole symmetry which in real fluids is absent due to finite size of the particles and the infinite repulsive potential at small distances \cite{book_lubenskycondmatt}. In particular the specific shape of the particles influences the equation of state \cite{hardsphereboublik_molphys1981}.

The Hamiltonian of the lattice gas is:
\begin{equation}\label{ham_latticegas}
  H = -J\suml_{
\left\langle\, ij \,\right\rangle
  } \, n_{i}\,n_{j} - \mu \,\suml_{i}\,n_{i}
\end{equation}
where $J$ is the energy of the site-site interaction of the nearest sites $i$ and $j$, $\mu$ is the chemical potential. This Hamiltonian is trivially isomorphic to the Hamiltonian of the Ising model of interacting spins $s_{i}$ via the transformation $n_{i} = \f{1+s_i}{2},\, s_i = \pm 1$ so that $n_{i} = 0,1$ whether the site is empty or occupied correspondingly. The order parameter is the probability of occupation $x = \left\langle\, n_i \,\right\rangle$ and serves as the analog of the density.

Indeed there are two specific elements which determine the structure of the liquid-gas part of the phase diagram. The first one is the coexistence line which is terminated by the second element  - the critical point (CP). The comparison of the phase diagram of the LG and the real fluid shows that they are topologically isomorphic. Therefore the one-to-one correspondence between the basic elements can be given (see \ref{fig_zeno_tri}). Moreover taking into account the linearities of the characteristic elements like diameter of the binodal and the Zeno-line such isomorphism can be chosen in the simplest form of the projective mapping since it conserves the incident and tangential  relations between the corresponding linear elements. Moreover the transformation of such class is unique because any projective transformation is determined by the correspondence between two triplets of
three nonconcurrent lines \cite{hartshorn_projgeom}. In our case we choose the following triplets the critical isotherms, the Zeno-lines and the Zeno-medians:
\begin{equation}\label{triples}
t_c = 1 \Leftrightarrow T = T_c\,,\quad x = 1/2 \Leftrightarrow \f{2n}{n_B}+\f{T}{T_B} = 1\,,\quad x = 1 \Leftrightarrow \f{n}{n_B}+\f{T}{T_B} = 1\,.
\end{equation}
As a sequence the binodal of the LG is mapped onto binodal of the real fluid. Thus the projective nature is based on
the fact that the ZL must be tangential to the extension of
the liquid branch of the binodal in low temperature region
stated in \cite{eos_zenoapfelbaum_jchemp2006}.

Basing on the consideration above we can find the
corresponding projective transformation between $(x,t)$ and
$(n,T)$ planes which maps these configurations onto each
other. This transformation has the following general form (see e.g. \cite{hartshorn_projgeom}):
\begin{equation}\label{ntx_proj}
  n =\f{a\,x+b\,t+c}{d\,x+e\,t+1}\,,\quad T = \f{h\,x+k\,t+l}{d\,x+e\,t+1}
\end{equation}
Taking into account the correspondence between the coordinate axes:
\[x=0\Leftrightarrow n=0 \quad \text{and}\quad  t=0 \Leftrightarrow T=0\]
e can put $c=l=h=b=0$. Finally we can write:
\begin{equation}
  n =\, n_B\,\f{x}{1+\alpha \,t}\,,\quad
  T =\, T_B\,\f{\alpha\, t}{1+\alpha\, t} \,, \label{proj_trans}
\end{equation}
where the parameter $\alpha$ can be derived from the correspondence between the critical points $x_c=1/2,\, t_c=1 \Leftrightarrow n=n_c,\, T=T_c$. Elementary algebra gives:
\[\alpha = \f{T_c}{T_B-T_c}\,.\]
It is easy to verify that the triples \eqref{triples} maps onto each other.
Obviously the value of $\alpha $ is determined by the interparticle potential.

Note that the Boyle temperature $T_B$ of the real fluid corresponds to the physically unreachable state with $t=\infty$ for the lattice gas. This reflects the fact that two particles of the LG can not occupy the same site because this is the prohibited configuration (occupation number $n_i$ is either 0 or 1). Additional argument comes from the consideration of the second virial coefficient. Indeed, $T_B$ is determined by the balance between repulsive and  attractive contributions into the second virial coefficient \cite{book_hansenmcdonald}:
\begin{equation}\label{tb}
T_B = \f{a}{b}\,,
\end{equation}
where
\[a = \pi \intl_{\sigma}^{+\infty}\,U(r)\,r^2\,dr\,,\quad b = \f{2\,\pi }{3}\,\sigma^3\,,\]
and $\sigma$ is the diameter of the hard core, $U(r)$ - the potential of the long range attractive part of the interaction. From this consideration it is clear that for point like particles $\sigma \to 0$ and $T_B \to \infty$ in accordance with \eqref{tb}.

Taking into account that the lattice gas CP $x_c = 1/2,\,t_c = 1$ maps onto the CP of real fluid $n_c, T_c$ from \eqref{proj_trans}  it is easy to get the value of the asymptotic density $n_B$:
\begin{equation}\label{nb}
  n_B = \f{2\,n_c}{1-\f{T_c}{T_B}}\,.
\end{equation}
This relation is the direct consequence of the proposed isomorphism and can be verified using the experimental data. In particular it is known that for many substances the approximate relation
\begin{equation}\label{tbtc}
T_B \approx 3\,T_c\,\,.
\end{equation}
is valid. Then from \eqref{nb} we get:
\begin{equation}\label{nbnc}
n_B \approx 3\,n_c\,\,.
\end{equation}
Below we show how \eqref{tbtc},\eqref{nbnc} follow directly from the \eqref{proj_trans} augmented by the scaling properties of the LG model. Also within such geometrical approach we can derive the linearity of the critical points line which was obtained in  \cite{eos_zenoapfelbaum_jchempb2008} by the consideration of the binodals of the liquids in reduced coordinates $T/T_B,\, n/n_B$.

Indeed different molecular fluids are described by the different scales for $T$ and $n$. For the corresponding lattice gas models it means that their parameters also scales in the same way. Basing on the structure of the LG Hamiltonian \eqref{ham_latticegas} we can state that the phase diagram of the LG is invariant under the scale transformation
\begin{equation}\label{scaletransform_xt}
t\to \lambda^2\,t\,,\,\, n_i\to \lambda^{-1}\,n_i
\end{equation}
as the LG temperature parameter $t$ is proportional to the interaction constant $J$. Note that this scale symmetry correlates exactly with the Lenard-Jones (LJ) potential where the leading interaction is of van der Waals nature $V(r) \propto  - \epsilon \,(\sigma /r)^6$. Omitting the influence of the repulsive part we get the same result since $T_c\sim \epsilon $ and $n_c \sim 1/\left\langle\, r \,\right\rangle_{c}^3$, where $\left\langle\, r \,\right\rangle_c$ is the average interparticle distance at the CP. From here it follows that $T_c\,n_c^2$ is invariant under the scale transformation similar to \eqref{scaletransform_xt}. Of course this is nothing but the PCS for the LJ fluids which can be easily extended to other interaction potentials which have the scale invariant properties \cite{book_prigozhisolut}.
\begin{figure}
\includegraphics[scale=0.75]{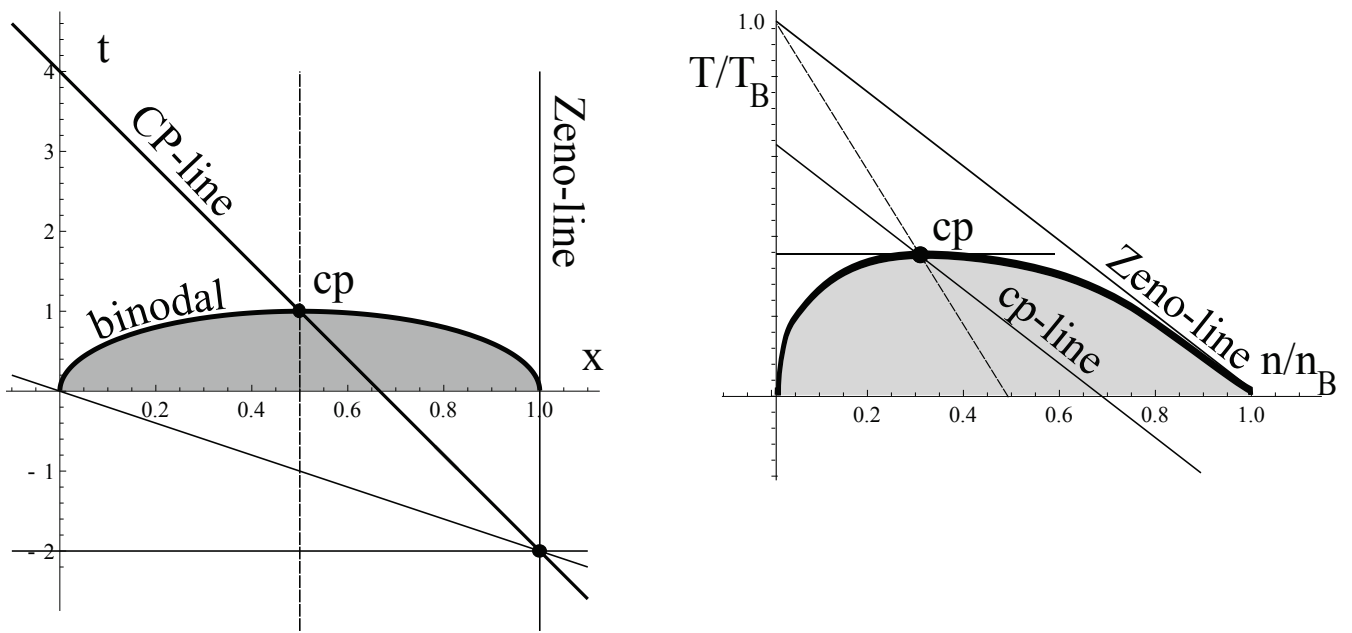}\\
  \caption{The Zeno and CP lines for LG and real fluids.}\label{fig_cpshift}
\end{figure}

The spatial scale in LG is determined by the state for which $x=1$ where the lattice of particles without the hole defects is formed. This corresponds to the asymptotic state of the infinite density $n=\infty$.  In the real fluid the cavities are always present. The relations \eqref{scaletransform_xt} together with the projective isomorphism \eqref{proj_trans} lead to the fact that the point $x=1, t=-2$ of intersection of the line
\begin{equation}\label{cp_shiftinf}
2x+t = 0
\end{equation}
with the Zeno-line $x=1$ of the LG determines the fixed point
of the scale transformation. Since the point $(1,-2)$ of
$(x,t)$-plane maps into infinite point, then \eqref{proj_trans} becomes as the following:
\begin{equation}\label{proj_trans_inf}
  n = n_B \, \f{x}{1+\f{t}{2}}\,,\quad T = \f{T_B}{2}\,\f{t}{1+\f{t}{2}}\,.
\end{equation}
From \eqref{proj_trans_inf} the empirical equalities:
\begin{equation}\label{empir_nctcnbtb}
  n_B = 3\,n_c\,,\,\,T_{B} = 3\,T_c\,\,,
\end{equation}
follow directly. They have been used above as the empirical
facts (see \eqref{nbnc}).
Under the infinitesimal transformation of  the scale
\eqref{scaletransform_xt} the locus of the lattice-gas CP
shifts along the line:
\begin{equation}\label{lgcp_line}
  t+6\,x  = 4\,,
\end{equation}
which connects the points $(1/2,1)$ and $(1,-2)$ on the LG
plane. Also, the line \eqref{lgcp_line} transforms into the
line of the critical points on $(n,T)$ plane:
\begin{equation}\label{cpline_my}
  \f{n_c}{n_B} + \f{T_c}{T_B} = \f{2}{3}\,,
\end{equation}
This situation is shown on \ref{fig_cpshift}.
Note that \eqref{cpline_my} perfectly corresponds with the
equation of the critical points line:
\begin{equation}\label{cpline_apfelbaum}
\f{n_{c}/n_B}{1-a}+ \f{T_{c}/T_B}{1-a} = 1\,,\quad a\approx 0.33\,,
\end{equation}
obtained in \cite{eos_zenoapfelbaum_jchempb2008} basing on the LJ model ``\textit{as the most consecutive one and the one closest to the real substances}``.

So we can conclude that the fact that the value of the parameter $\alpha$ is close to $1/2$ reflects the scale invariance property of the ``$r^{-6}$`` attractive part of the LJ potential similar to that for the LG Hamiltonian. For the real substances the parameter $\alpha$ depends on the repulsive part also so that $\alpha =1/2$ is no more valid. In such a case the general form of the CP line is:
\begin{equation}\label{cplinealpha}
  \f{n_c}{n_B}+  \f{T_c}{T_B} = \f{1+2\,\alpha }{2+2\,\alpha }\,.
\end{equation}
Nevertheless for many substances the value of $\alpha$ is close to $1/2$ and correspondingly the estimates \eqref{empir_nctcnbtb} are quite good. This allows to give the classical Principle of Corresponding States new formulation using the proposed approach.

\section{Discussion}
We have shown that all the empirical relations about the
linearity of the Zeno-line and the rectilinear diameter can
casted into geometrical form. The proposed geometrical picture allows to get the unifying
view on the empirical relations obtained in
\cite{eos_zenobenamotz_isrchemphysj1990,eos_zenoapfelbaum_jcp2004,
eos_zenoapfelbaum_jchemp2006,eos_zenoapfelbaum_jchemp2008}. Such geometrical interpretation is based on the essential
similarity of the LJ liquid isomorphism and scaling symmetry
for the Hamiltonians of the LG and the LJ fluids.

In conclusion we discuss the range of validity of the proposed isomorphism. Of course the proposed isomorphism is approximate just like the PCS.
Nevertheless it provides the useful zeroth approximation for the estimates of the corresponding properties of the real
substances \cite{liqmetals_zenoapfelbaum2_cpl2009}.

The range
of the validity for such simple projective isomorphism is based on the conservation of the essential features of the LG. They
are (a) independence of the interaction on the thermodynamical state ; (b) the absence of the formation of the additional
structures in liquids, e.g. polymerization, H-bond network etc. From this point of view it is expected that the linear
character of both the Zeno-line and the rectilinear diameter
rely on the LJ character of the interparticle interactions.
This correlates with the known facts that the RDL and the
linearity of the Zeno-line are violated in liquid metals metals \cite{liqmetals_singdiamhensel_prl1985,*liqmetals_diamliqmetals_hensel_jphys1996} and water \cite{eos_zeno_jphyschemb2000}. Also the triple point which exists in real fluids and is absent for LG leads to the
deviation of the isomorphism from the simple form of the
projective transformation \eqref{proj_trans}. Indeed the Zeno-line is defined as the tangent to the binodal in the physical inaccessible region of $T\to 0$ and must correctly describe the low-temperature part of the binodal \cite{eos_zenoapfelbaum_jchemp2006, zenoline_potentials_jcp2009}. It would be interesting to elaborate the extension of the proposed approach taking into account the existence of the triple point which bounds the coexistence line at finite densities.

The growth of the fluctuations in the vicinity of the CP is
responsible for the deviation from the linearities which
specified by the projective mapping. The latter is nonlinear
and therefore $\left\langle\, f(n_i) \,\right\rangle \ne f(\left\langle\, n_i \,\right\rangle)
$, where $
\left\langle\, \ldots \,\right\rangle
$ is the statistical average. In particular, in the vicinity of the CP the rectilinear
diameter deviates from the simple linear law and shows
anomalies \cite{crit_diamermin1_prl1971,*crit_diamermin2_prl1971,*crit_rehrmermin_pra1973}. Because of this the Zeno-median does not coincide with the
diameter and approaches it only asymptotically far away from
the CP \cite{eos_zenobenamotz_isrchemphysj1990,eos_zenoapfelbaum_jchemp2006}.

Such deviations could give the support to search the
isomorphism on the basis of the initial statistical local
fields like in the approach proposed
\cite{crit_can_kulinskii_jmolliq2003,*crit_can_diamsing_kulimalo_physa2009} rather than the nonlinear transformation of the thermodynamic
averages only
\cite{crit_yydiamfisherorkoulas_prl2000,*crit_fishmixdiam1_pre2003,
*crit_aniswangasymmetry_pre2007}.

\begin{acknowledgements}
The author cordially thank Prof. Malomuzh for the discussion of obtained results.
\end{acknowledgements}


\begin{thebibliography}{30}
\expandafter\ifx\csname natexlab\endcsname\relax\def\natexlab#1{#1}\fi
\expandafter\ifx\csname bibnamefont\endcsname\relax
  \def\bibnamefont#1{#1}\fi
\expandafter\ifx\csname bibfnamefont\endcsname\relax
  \def\bibfnamefont#1{#1}\fi
\expandafter\ifx\csname citenamefont\endcsname\relax
  \def\citenamefont#1{#1}\fi
\expandafter\ifx\csname url\endcsname\relax
  \def\url#1{\texttt{#1}}\fi
\expandafter\ifx\csname urlprefix\endcsname\relax\def\urlprefix{URL }\fi
\providecommand{\bibinfo}[2]{#2}
\providecommand{\eprint}[2][]{\url{#2}}

\bibitem[{\citenamefont{van~der Waals}(1873)}]{vdW_thesis}
\bibinfo{author}{\bibfnamefont{J.~D.} \bibnamefont{van~der Waals}}, Ph.D.
  thesis, \bibinfo{school}{Lieden University}, \bibinfo{address}{Leiden}
  (\bibinfo{year}{1873}).

\bibitem[{\citenamefont{Prigogine}(1957)}]{book_prigozhisolut}
\bibinfo{author}{\bibfnamefont{I.}~\bibnamefont{Prigogine}},
  \emph{\bibinfo{title}{The Molecular Theory of Solutions}}
  (\bibinfo{publisher}{Interscience Publihers}, \bibinfo{address}{NY},
  \bibinfo{year}{1957}).

\bibitem[{\citenamefont{Kulinskii et~al.}(2009)\citenamefont{Kulinskii,
  Malomuzh, and Matvejchuk}}]{crit_dimers_noblepcs_physica2009}
\bibinfo{author}{\bibfnamefont{V.~L.} \bibnamefont{Kulinskii}},
  \bibinfo{author}{\bibfnamefont{N.~P.} \bibnamefont{Malomuzh}},
  \bibnamefont{and} \bibinfo{author}{\bibfnamefont{O.~I.}
  \bibnamefont{Matvejchuk}}, \bibinfo{journal}{Physica A: Statistical Mechanics
  and its Applications} \textbf{\bibinfo{volume}{388}}, \bibinfo{pages}{4560}
  (\bibinfo{year}{2009}), ISSN \bibinfo{issn}{03784371},
  \urlprefix\url{http://dx.doi.org/10.1016/j.physa.2009.07.011}.

\bibitem[{\citenamefont{Cailletet and Mathias}(1886)}]{crit_diam0}
\bibinfo{author}{\bibfnamefont{L.}~\bibnamefont{Cailletet}} \bibnamefont{and}
  \bibinfo{author}{\bibfnamefont{E.}~\bibnamefont{Mathias}},
  \bibinfo{journal}{Scanc. Acad. Sci. Compt. Rend. Hebd., Paris}
  \textbf{\bibinfo{volume}{102}}, \bibinfo{pages}{1202} (\bibinfo{year}{1886}).

\bibitem[{\citenamefont{Young}(1900)}]{crit_diam1_young_philmag1900}
\bibinfo{author}{\bibfnamefont{S.}~\bibnamefont{Young}},
  \bibinfo{journal}{Phil. Mag.} \textbf{\bibinfo{volume}{50}},
  \bibinfo{pages}{291} (\bibinfo{year}{1900}).

\bibitem[{\citenamefont{Patashinskii and Pokrovsky}(1979)}]{book_patpokr}
\bibinfo{author}{\bibfnamefont{A.~Z.} \bibnamefont{Patashinskii}}
  \bibnamefont{and} \bibinfo{author}{\bibfnamefont{V.~L.}
  \bibnamefont{Pokrovsky}}, \emph{\bibinfo{title}{Fluctuation theory of
  critical phenomena}} (\bibinfo{publisher}{Pergamon},
  \bibinfo{address}{Oxford}, \bibinfo{year}{1979}).

\bibitem[{\citenamefont{Weiss and
  Schr\"{o}er}(2008)}]{crit_diam_schroerweiss_statphys2008}
\bibinfo{author}{\bibfnamefont{V.~C.} \bibnamefont{Weiss}} \bibnamefont{and}
  \bibinfo{author}{\bibfnamefont{W.}~\bibnamefont{Schr\"{o}er}},
  \bibinfo{journal}{J. Stat. Mech.} \textbf{\bibinfo{volume}{2008}},
  \bibinfo{pages}{P04020 (26pp)} (\bibinfo{year}{2008}),
  \urlprefix\url{http://stacks.iop.org/1742-5468/2008/P04020}.

\bibitem[{\citenamefont{Batschinski}(1906)}]{eos_zenobatschinski_annphys1906}
\bibinfo{author}{\bibfnamefont{A.}~\bibnamefont{Batschinski}},
  \bibinfo{journal}{Ann. Phys.} \textbf{\bibinfo{volume}{19}},
  \bibinfo{pages}{307} (\bibinfo{year}{1906}).

\bibitem[{\citenamefont{Ben-Amotz and
  Herschbach}(1990)}]{eos_zenobenamotz_isrchemphysj1990}
\bibinfo{author}{\bibfnamefont{D.}~\bibnamefont{Ben-Amotz}} \bibnamefont{and}
  \bibinfo{author}{\bibfnamefont{D.~R.} \bibnamefont{Herschbach}},
  \bibinfo{journal}{Isr. J. Chem.} \textbf{\bibinfo{volume}{30}},
  \bibinfo{pages}{59} (\bibinfo{year}{1990}).

\bibitem[{\citenamefont{Apfelbaum et~al.}(2008)\citenamefont{Apfelbaum,
  Vorob'ev, and Martynov}}]{eos_zenoapfelbaum_jchemp2008}
\bibinfo{author}{\bibfnamefont{E.~M.} \bibnamefont{Apfelbaum}},
  \bibinfo{author}{\bibfnamefont{V.~S.} \bibnamefont{Vorob'ev}},
  \bibnamefont{and} \bibinfo{author}{\bibfnamefont{G.~A.}
  \bibnamefont{Martynov}}, \bibinfo{journal}{J. Phys. Chem. A}
  \textbf{\bibinfo{volume}{112}}, \bibinfo{pages}{6042} (\bibinfo{year}{2008}).

\bibitem[{\citenamefont{Apfelbaum et~al.}(2004)\citenamefont{Apfelbaum,
  Vorob'ev, and Martynov}}]{eos_zenoapfelbaum_jcp2004}
\bibinfo{author}{\bibfnamefont{E.~M.} \bibnamefont{Apfelbaum}},
  \bibinfo{author}{\bibfnamefont{V.~S.} \bibnamefont{Vorob'ev}},
  \bibnamefont{and} \bibinfo{author}{\bibfnamefont{G.~A.}
  \bibnamefont{Martynov}}, \bibinfo{journal}{J. Phys. Chem. A}
  \textbf{\bibinfo{volume}{108}}, \bibinfo{pages}{10381}
  (\bibinfo{year}{2004}), \urlprefix\url{http://dx.doi.org/10.1021/jp046417z}.

\bibitem[{\citenamefont{Apfelbaum et~al.}(2006)\citenamefont{Apfelbaum,
  Vorob'ev, and Martynov}}]{eos_zenoapfelbaum_jchemp2006}
\bibinfo{author}{\bibfnamefont{E.~M.} \bibnamefont{Apfelbaum}},
  \bibinfo{author}{\bibfnamefont{V.~S.} \bibnamefont{Vorob'ev}},
  \bibnamefont{and} \bibinfo{author}{\bibfnamefont{G.~A.}
  \bibnamefont{Martynov}}, \bibinfo{journal}{J. Phys. Chem. B}
  \textbf{\bibinfo{volume}{110}}, \bibinfo{pages}{8474} (\bibinfo{year}{2006}).

\bibitem[{\citenamefont{Apfelbaum and
  Vorob'ev}(2009{\natexlab{a}})}]{liqmetals_zenoapfelbaum2_cpl2009}
\bibinfo{author}{\bibfnamefont{E.~M.} \bibnamefont{Apfelbaum}}
  \bibnamefont{and} \bibinfo{author}{\bibfnamefont{V.~S.}
  \bibnamefont{Vorob'ev}}, \bibinfo{journal}{Chem. Phys. Lett.}
  \textbf{\bibinfo{volume}{467}}, \bibinfo{pages}{318}
  (\bibinfo{year}{2009}{\natexlab{a}}).

\bibitem[{\citenamefont{Chaikin and Lubensky}(2000)}]{book_lubenskycondmatt}
\bibinfo{author}{\bibfnamefont{P.~M.} \bibnamefont{Chaikin}} \bibnamefont{and}
  \bibinfo{author}{\bibfnamefont{T.~C.} \bibnamefont{Lubensky}},
  \emph{\bibinfo{title}{Principles of Condensed Matter Physics}}
  (\bibinfo{publisher}{Cambridge University Press}, \bibinfo{year}{2000}),
  \bibinfo{edition}{new edition} ed., ISBN \bibinfo{isbn}{0521794501},
  \urlprefix\url{http://www.amazon.ca/exec/obidos/redirect?tag=citeulike09-20\%
&amp;path=ASIN/0521794501}.

\bibitem[{\citenamefont{Boublik}(1981)}]{hardsphereboublik_molphys1981}
\bibinfo{author}{\bibfnamefont{T.}~\bibnamefont{Boublik}},
  \bibinfo{journal}{Mol. Phys.} \textbf{\bibinfo{volume}{42}},
  \bibinfo{pages}{209} (\bibinfo{year}{1981}),
  \urlprefix\url{http://dx.doi.org/10.1080/00268978100100161}.

\bibitem[{\citenamefont{Hartshorne}(1968)}]{hartshorn_projgeom}
\bibinfo{author}{\bibfnamefont{R.}~\bibnamefont{Hartshorne}},
  \emph{\bibinfo{title}{Foundations of Projective Geometry}}
  (\bibinfo{publisher}{Addison Wesley Publishing Company},
  \bibinfo{year}{1968}), ISBN \bibinfo{isbn}{0805337571},
  \urlprefix\url{http://www.amazon.com/exec/obidos/redirect?tag=citeulike07-20%
&path=ASIN/0805337571}.

\bibitem[{\citenamefont{Hansen and Mcdonald}(2006)}]{book_hansenmcdonald}
\bibinfo{author}{\bibfnamefont{J.-P.} \bibnamefont{Hansen}} \bibnamefont{and}
  \bibinfo{author}{\bibfnamefont{I.~R.} \bibnamefont{Mcdonald}},
  \emph{\bibinfo{title}{Theory of Simple Liquids, Third Edition}}
  (\bibinfo{publisher}{{Academic Press}}, \bibinfo{year}{2006}), ISBN
  \bibinfo{isbn}{0123705355}.

\bibitem[{\citenamefont{Apfelbaum and
  Vorob'ev}(2008)}]{eos_zenoapfelbaum_jchempb2008}
\bibinfo{author}{\bibfnamefont{E.~M.} \bibnamefont{Apfelbaum}}
  \bibnamefont{and} \bibinfo{author}{\bibfnamefont{V.~S.}
  \bibnamefont{Vorob'ev}}, \bibinfo{journal}{J. Phys. Chem B.}
  \textbf{\bibinfo{volume}{112}}, \bibinfo{pages}{13064}
  (\bibinfo{year}{2008}).

\bibitem[{\citenamefont{J\"{u}ngst et~al.}(1985)\citenamefont{J\"{u}ngst,
  Knuth, and Hensel}}]{liqmetals_singdiamhensel_prl1985}
\bibinfo{author}{\bibfnamefont{S.}~\bibnamefont{J\"{u}ngst}},
  \bibinfo{author}{\bibfnamefont{B.}~\bibnamefont{Knuth}}, \bibnamefont{and}
  \bibinfo{author}{\bibfnamefont{F.}~\bibnamefont{Hensel}},
  \bibinfo{journal}{Phys. Rev. Lett.} \textbf{\bibinfo{volume}{55}},
  \bibinfo{pages}{2160} (\bibinfo{year}{1985}).

\bibitem[{\citenamefont{Ross and
  Hensel}(1996)}]{liqmetals_diamliqmetals_hensel_jphys1996}
\bibinfo{author}{\bibfnamefont{M.}~\bibnamefont{Ross}} \bibnamefont{and}
  \bibinfo{author}{\bibfnamefont{F.}~\bibnamefont{Hensel}},
  \bibinfo{journal}{Journal of Physics: Condensed Matter}
  \textbf{\bibinfo{volume}{8}}, \bibinfo{pages}{1909} (\bibinfo{year}{1996}),
  \urlprefix\url{http://dx.doi.org_10.1088_0953-8984_8_12_006}.

\bibitem[{\citenamefont{Kutney et~al.}(2000)\citenamefont{Kutney, Reagan,
  Smith, Tester, and Herschbach}}]{eos_zeno_jphyschemb2000}
\bibinfo{author}{\bibfnamefont{M.~C.} \bibnamefont{Kutney}},
  \bibinfo{author}{\bibfnamefont{M.~T.} \bibnamefont{Reagan}},
  \bibinfo{author}{\bibfnamefont{K.~A.} \bibnamefont{Smith}},
  \bibinfo{author}{\bibfnamefont{J.~W.} \bibnamefont{Tester}},
  \bibnamefont{and} \bibinfo{author}{\bibfnamefont{D.~R.}
  \bibnamefont{Herschbach}}, \bibinfo{journal}{J. Phys. Chem. B}
  \textbf{\bibinfo{volume}{104}}, \bibinfo{pages}{9513} (\bibinfo{year}{2000}).

\bibitem[{\citenamefont{Apfelbaum and
  Vorob'ev}(2009{\natexlab{b}})}]{zenoline_potentials_jcp2009}
\bibinfo{author}{\bibfnamefont{E.~M.} \bibnamefont{Apfelbaum}}
  \bibnamefont{and} \bibinfo{author}{\bibfnamefont{V.~S.}
  \bibnamefont{Vorob'ev}}, \bibinfo{journal}{J. Chem. Phys.}
  \textbf{\bibinfo{volume}{130}}, \bibinfo{eid}{214111}
  (pages~\bibinfo{numpages}{10}) (\bibinfo{year}{2009}{\natexlab{b}}),
  \urlprefix\url{http://link.aip.org/link/?JCP/130/214111/1}.

\bibitem[{\citenamefont{Mermin}(1971{\natexlab{a}})}]{crit_diamermin1_prl1971}
\bibinfo{author}{\bibfnamefont{N.~D.} \bibnamefont{Mermin}},
  \bibinfo{journal}{Phys. Rev. Lett.} \textbf{\bibinfo{volume}{26}},
  \bibinfo{pages}{169} (\bibinfo{year}{1971}{\natexlab{a}}).

\bibitem[{\citenamefont{Mermin}(1971{\natexlab{b}})}]{crit_diamermin2_prl1971}
\bibinfo{author}{\bibfnamefont{N.~D.} \bibnamefont{Mermin}},
  \bibinfo{journal}{Phys. Rev. Lett.} \textbf{\bibinfo{volume}{26}},
  \bibinfo{pages}{957} (\bibinfo{year}{1971}{\natexlab{b}}).

\bibitem[{\citenamefont{Rehr and Mermin}(1973)}]{crit_rehrmermin_pra1973}
\bibinfo{author}{\bibfnamefont{J.~J.} \bibnamefont{Rehr}} \bibnamefont{and}
  \bibinfo{author}{\bibfnamefont{N.~D.} \bibnamefont{Mermin}},
  \bibinfo{journal}{Physical Review A} \textbf{\bibinfo{volume}{8}},
  \bibinfo{pages}{472} (\bibinfo{year}{1973}),
  \urlprefix\url{http://dx.doi.org/10.1103/PhysRevA.8.472}.

\bibitem[{\citenamefont{Kulinskii}(2003)}]{crit_can_kulinskii_jmolliq2003}
\bibinfo{author}{\bibfnamefont{V.}~\bibnamefont{Kulinskii}},
  \bibinfo{journal}{J. Mol. Liq.} \textbf{\bibinfo{volume}{105}},
  \bibinfo{pages}{273} (\bibinfo{year}{2003}), ISSN \bibinfo{issn}{01677322},
  \urlprefix\url{http://dx.doi.org/10.1016/S0167-7322(03)00067-9}.

\bibitem[{\citenamefont{Kulinskii and
  Malomuzh}(2009)}]{crit_can_diamsing_kulimalo_physa2009}
\bibinfo{author}{\bibfnamefont{V.}~\bibnamefont{Kulinskii}} \bibnamefont{and}
  \bibinfo{author}{\bibfnamefont{N.}~\bibnamefont{Malomuzh}},
  \bibinfo{journal}{Physica A: Statistical Mechanics and its Applications}
  \textbf{\bibinfo{volume}{388}}, \bibinfo{pages}{621} (\bibinfo{year}{2009}),
  ISSN \bibinfo{issn}{03784371},
  \urlprefix\url{http://dx.doi.org/10.1016/j.physa.2008.11.014}.

\bibitem[{\citenamefont{Fisher and
  Orkoulas}(2000)}]{crit_yydiamfisherorkoulas_prl2000}
\bibinfo{author}{\bibfnamefont{M.~E.} \bibnamefont{Fisher}} \bibnamefont{and}
  \bibinfo{author}{\bibfnamefont{G.}~\bibnamefont{Orkoulas}},
  \bibinfo{journal}{Phys. Rev. Lett.} \textbf{\bibinfo{volume}{85}},
  \bibinfo{pages}{696} (\bibinfo{year}{2000}).

\bibitem[{\citenamefont{Kim et~al.}(2003)\citenamefont{Kim, Fisher, and
  Orkoulas}}]{crit_fishmixdiam1_pre2003}
\bibinfo{author}{\bibfnamefont{Y.~C.} \bibnamefont{Kim}},
  \bibinfo{author}{\bibfnamefont{M.~E.} \bibnamefont{Fisher}},
  \bibnamefont{and} \bibinfo{author}{\bibfnamefont{G.}~\bibnamefont{Orkoulas}},
  \bibinfo{journal}{Phys. Rev. E} \textbf{\bibinfo{volume}{67}},
  \bibinfo{pages}{061506} (\bibinfo{year}{2003}).

\bibitem[{\citenamefont{Wang and
  Anisimov}(2007)}]{crit_aniswangasymmetry_pre2007}
\bibinfo{author}{\bibfnamefont{J.}~\bibnamefont{Wang}} \bibnamefont{and}
  \bibinfo{author}{\bibfnamefont{M.~A.} \bibnamefont{Anisimov}},
  \bibinfo{journal}{Physical Review E (Statistical, Nonlinear, and Soft Matter
  Physics)} \textbf{\bibinfo{volume}{75}}, \bibinfo{eid}{051107}
  (pages~\bibinfo{numpages}{19}) (\bibinfo{year}{2007}),
  \urlprefix\url{http://link.aps.org/abstract_prev75/e051107}.

\end{thebibliography}
\end{document}